\documentclass[fleqn,twoside]{article}
\usepackage{espcrc1}

\usepackage{graphicx}
\usepackage[figuresright]{rotating}

\newcommand{\AmS}{{\protect\the\textfont2
  A\kern-.1667em\lower.5ex\hbox{M}\kern-.125emS}}

\def\dm2{\Delta m^2}
\def\sq2{sin^2(2\Theta)}

\title{\bf Cosmic Ray composition around the knee from EAS electromagnetic and
muon data}

\author{The EAS-TOP and MACRO Collaboration
\thanks{The complete list of authors will be reported in the summary of the 
TAUP2001 Proceedings volume.}}       
\begin{document}

\maketitle

\begin{abstract}
\vspace{1pc}
We report
the analysis of the N$_e$-N$_\mu$  coincident events collected by
the MACRO/EAS-TOP collaboration at the Gran Sasso Laboratories.
The result points to a primary composition becoming heavier around the
knee of the primary spectrum (in the energy region $10^{15}-10^{16}$ eV).
The result is in very good agreement with the measurements
of EAS-TOP alone at the surface, wich detects muons with energy
$E_{\mu} > 1$ GeV and uses the same (QGSJET) interaction model.
\end{abstract}


\section{Introduction and the detectors}
The study of the primary composition in the Extensive Air Shower energy region 
requires the use of different observables in order to cross check the
information and reduce the dependence on the interaction model and
propagation codes used. 
At the National Gran Sasso Laboratories a program of exploiting the
surface  shower size measurements from  EAS-TOP (2005 m a.s.l.), and the
high energy muon measurements ($E_{\mu}^{th} = 1.3 $ TeV) performed in the
deep underground laboratories (MACRO) has been developed.
Such muons in fact originate from the decays of mesons produced in the
first interactions in the atmosphere and from a quite different rapidity region
than the GeV muons usually used for such analysis ($x_{F}> 0.1$ or $ 0.2$).

The two experiments operated in coincidence for a live time of
$\Delta T =$ 960.12 days 
between 1990 and 2000.
The number of coincident events collected with the two detectors operating
in the final configuration amounts to 28160, of which 3752
have shower cores inside the edges of the array ("internal events")
and shower size $N_e > 2.10^5$, and 409 have $N_e > 10^{5.92}$, i.e.
above the observed knee at the corresponding zenith angle.
We present here an analysis of the full data set in terms of the QGSJET 
interaction model as implemented in CORSIKA. 
For a comparison of different interaction models see \cite{as1}.

The EAS-TOP array is located at Campo Imperatore (2005 m a.s.l., $ \approx
30^o$ 
with respect to the vertical of the underground Gran Sasso Laboratories,
corresponding to 930 gr cm$^{-2}$ atmospheric depth).
Its e.m. detector (to which we are mainly interested in the present analysis)
is built of 35 scintillator modules 10 m$^2$ each, including
an area A $\approx$ 10$^5$ m$^2$.
The array is fully efficient for $N_e > 10^5$.
Its reconstruction capabilities of the EAS parameters 
(for internal events) are:
${{\Delta N_e} \over N_e} \approx 10 \%$ above $N_e \approx 10^5$,
and $\Delta \theta \approx 0.9^o$ for the EAS arrival direction.
The array and the reconstruction procedures are fully described in \cite{as2}.

MACRO, in the underground Gran Sasso Laboratory at 963 m a.s.l., 
with 3100 m w.e.
of minimum rock overburden, is a large area multi-purpose apparatus designed to
 detect penetrating cosmic radiation. The lower part of the MACRO detector has 
dimensions $76.6 \times 12 \times 4.8$ m$^3$.
A detailed description of the apparatus can be found in \cite{as21}.

In this work we consider only muon tracks which have at least
4 aligned hits in both views of the horizontal streamer tube planes
over the 10 layers composing the whole detector.
The standard reconstruction procedure of MACRO \cite{as3} 
has been used.

The two experiments are separated by a thickness of rock ranging from 1100 up 
to 1300 m, depending on the angle.

The muon energy threshold at the surface for muons reaching the MACRO depth
ranges from $E_{\mu}^{th} = 1.3$ TeV to $E_{\mu}^{th} = 1.8$ TeV inside 
the effective area of EAS-TOP.   
Event coincidence is established off-line, using the absolute time given by
a GPS system with an accuracy better than 1 $\mu s$.
 
Independent analyses of the two arrays are in \cite{as4} and \cite{as5}.

\section{Analysis and results}
The analysis technique has to be adapted 
to the specific trigger requirements (both surface and underground 
detectors fired) with defined acceptance area for the EAS array
(internal events), but undefined for the underground one.
Therefore, the main experimental feature to be considered
is the muon multiplicity distribution in different intervals
of shower sizes. We have chosen six intervals of shower sizes around the
knee: \\ 
5.20 $< Log_{10}(Ne)\leq$ 5.31;
5.31 $< Log_{10}(Ne)\leq$ 5.61; 
5.61 $< Log_{10}(Ne)\leq$ 5.92; 
5.92 $< Log_{10}(Ne)\leq$ 6.15; 
6.15 $< Log_{10}(Ne)\leq$ 6.35 and
6.35 $< Log_{10}(Ne)\leq$ 6.7.

Within each size bin the muon multiplicity distribution has been 
fitted with a superposition of pure $p$ and $Fe$ components, or light $(L)$ and
heavy $(H)$ admixtures containing equal fractions of $p$ and $He$ or $Mg$ and 
$Fe$ respectively.  
All spectra in the simulation have slope $\gamma = 2.62$.

The fit has been performed in the quoted six windows
by minimizing the following expression for each
multiplicity distribution:
\begin{equation}
\chi^2 = \sum_i{\frac{(N^{exp}_i - p_1 N^{p}_i - p_2 N^{Fe}_i)^2
}{\sigma_{i,exp}^2}} 
\end{equation}
 where $N^{exp}_i$ in the number of observed events in the $i$-th bin
of multiplicity, $N^p$ ($N^L$) and $N^{Fe}$ ($N^H$) are the number of 
simulated events
in the same $i$-th multiplicity bin from the $p$ ($L$) and $Fe$ ($H$) 
components, respectively; 
$p_1$ and $p_2$ are the parameters (to be fitted) defining the
fraction of each mass component contributing to  the same multiplicity bin.
The results of the fits have been normalized to reproduce the observed
number of coincident events in each size bin.

\begin{figure}[!htb]
\begin{center}
\includegraphics[width=10.0cm]{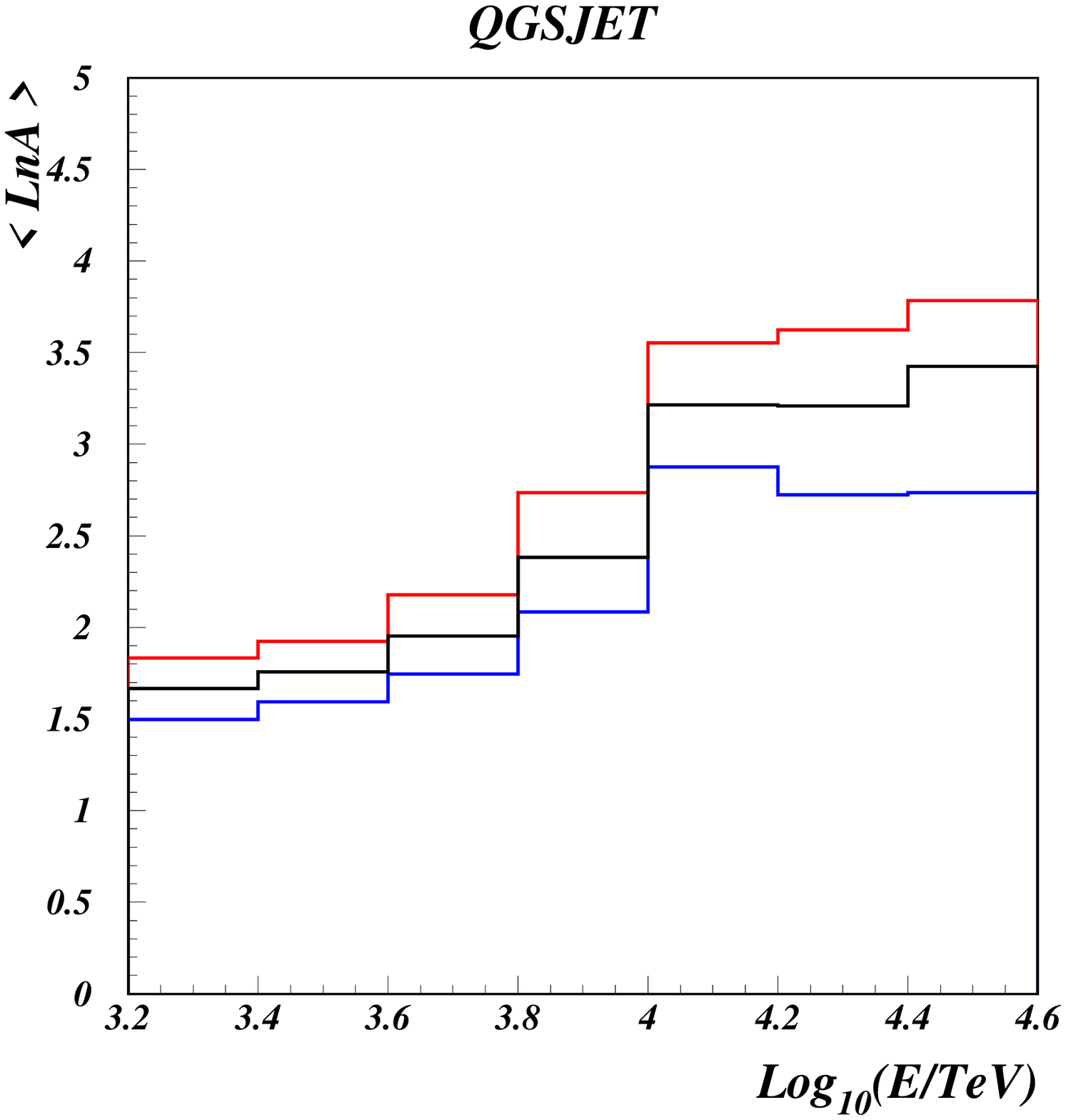}
\includegraphics[width=10.0cm]{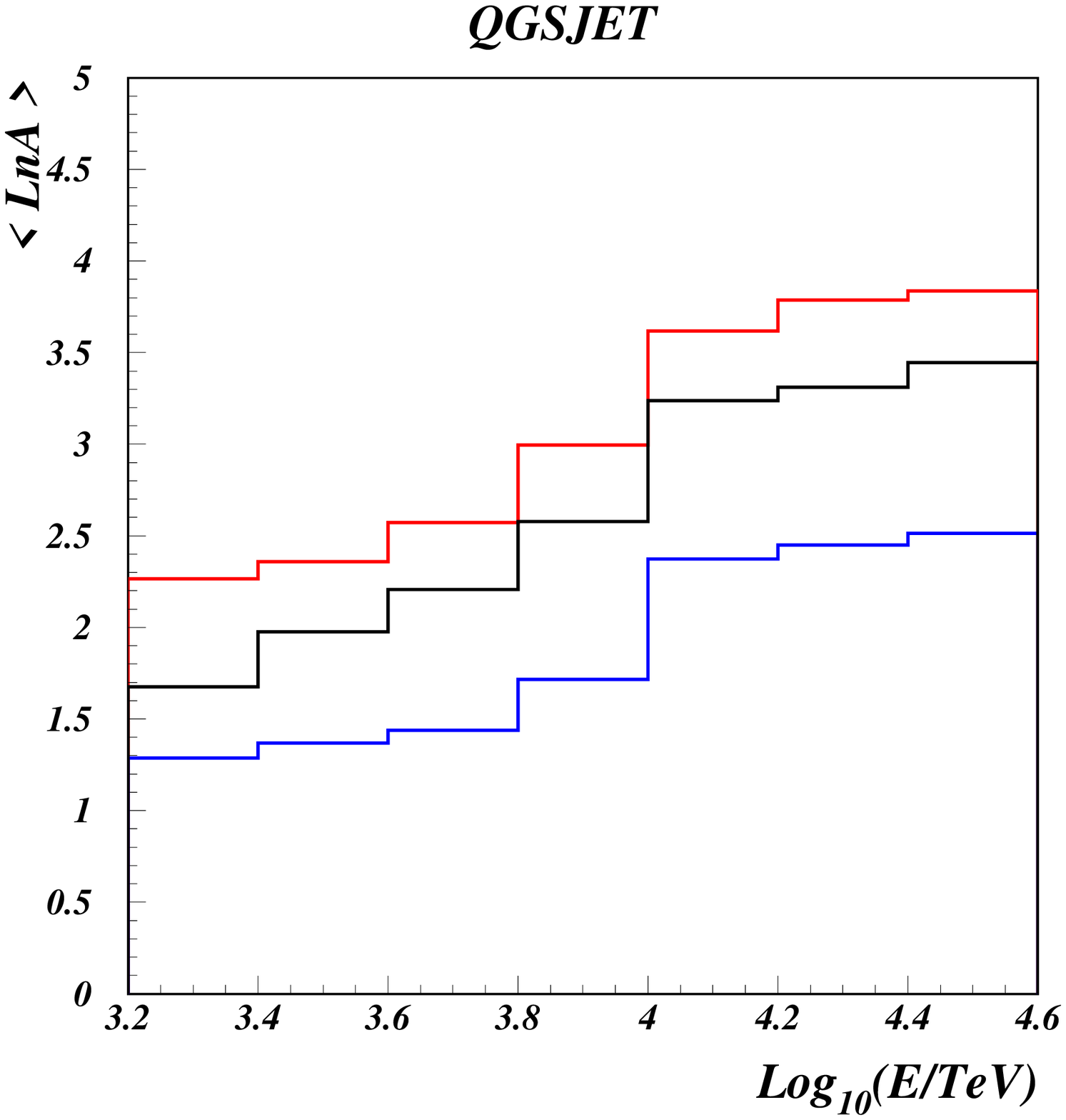}
\caption{$<log A>$ vs energy for $p/Fe$ (upper) and $Light/Heavy$ 
(lower) compositions.
The two lines include the statistical uncertainities.
\label{fig:aval} }
\end{center}
\end{figure}

To infer the mass composition 
evolution as a function of energy,
for each size bin we take from the
simulation the $log(E)$ distributions of contributing mass groups
weighted by the parameters $p_1$ and $p_2$ and with weights
$w_k$ representing the relative efficiency to trigger the underground 
apparatus.
The resulting distributions from different size bins are summed together,
and so we eventually obtain the simulated energy spectra of the two basic
components that reproduce the experimental data.
\begin{figure}[htb]
\begin{center}
\includegraphics[width=7cm]{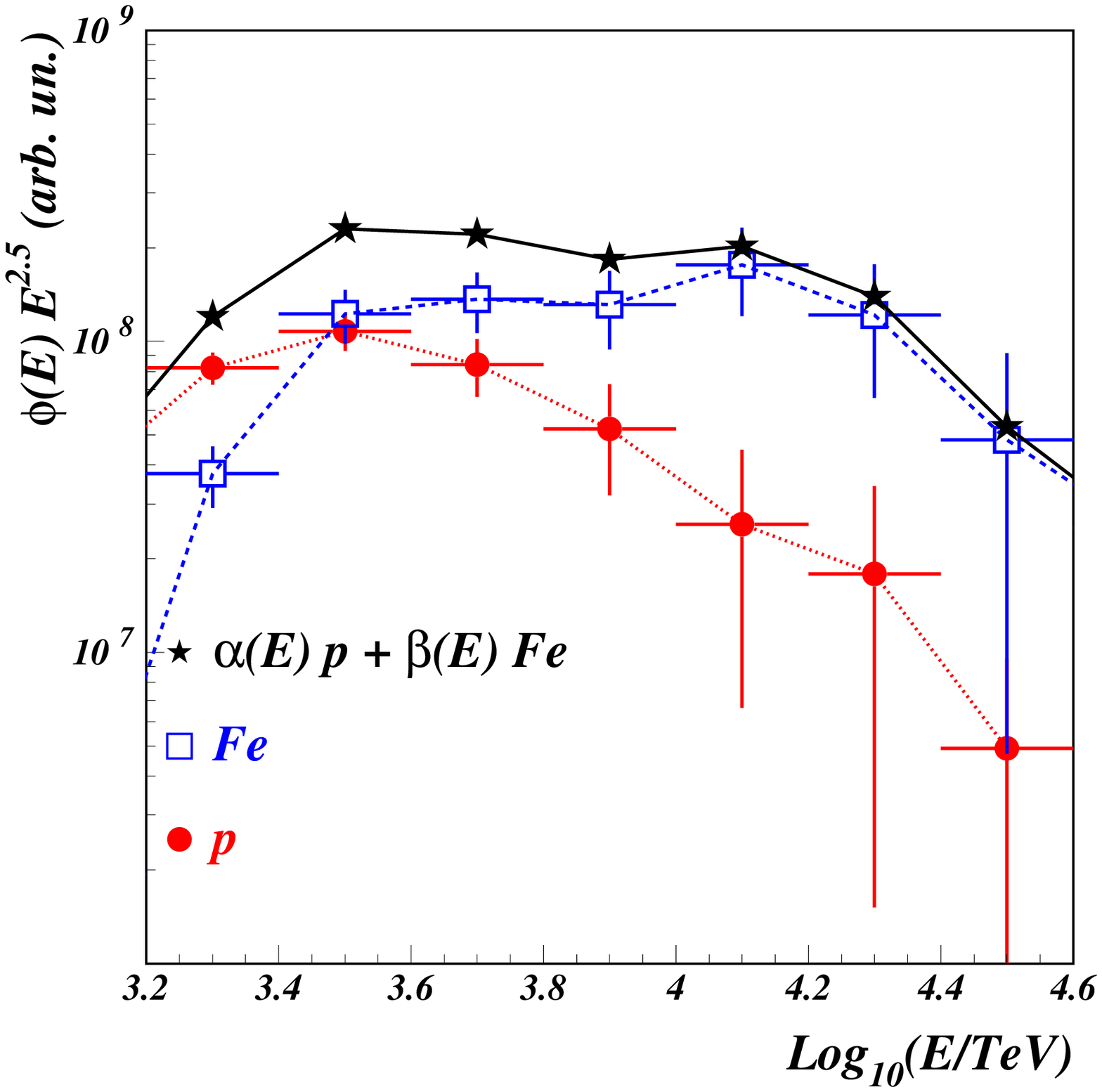}
\includegraphics[width=7cm]{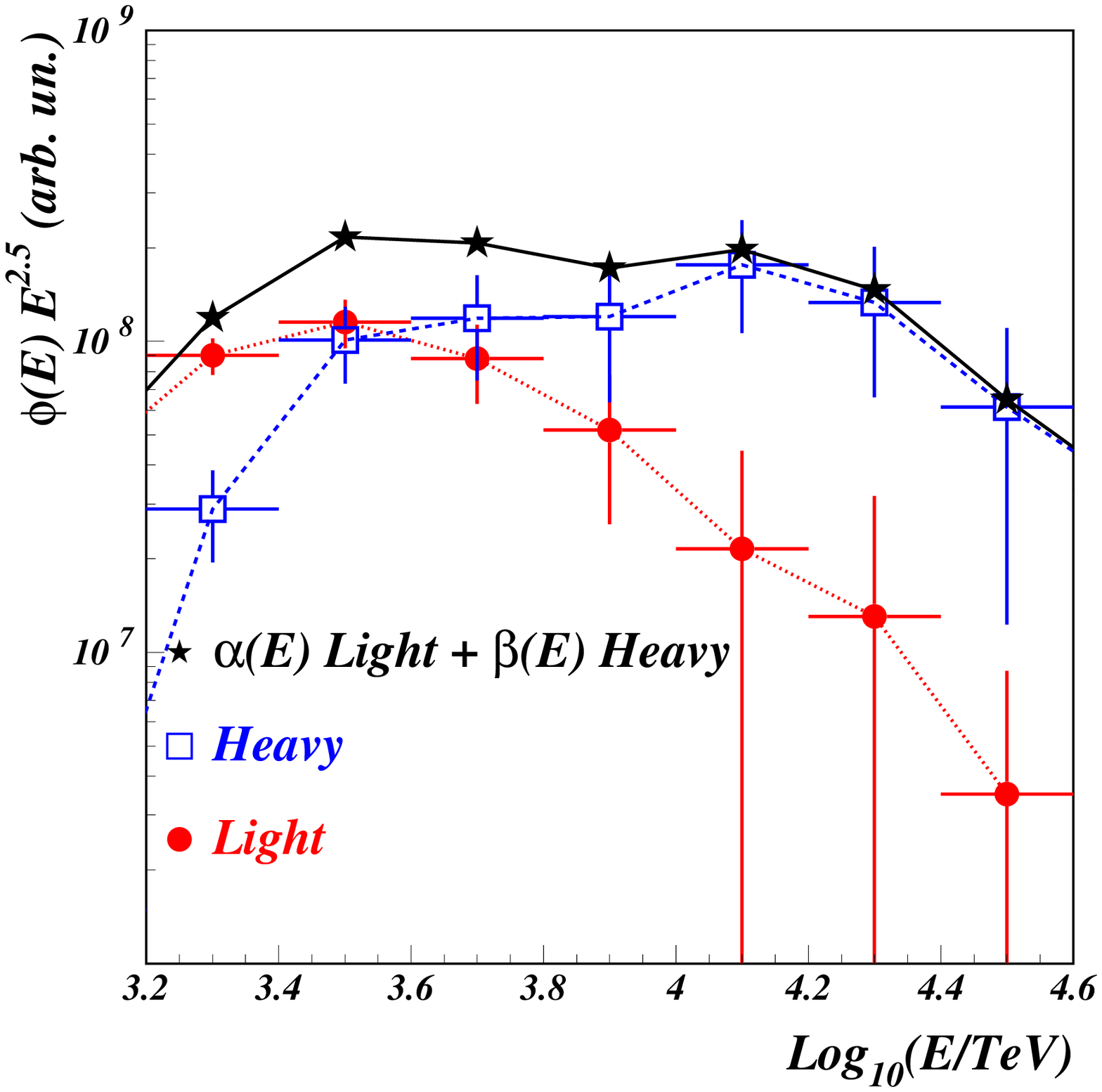}
\caption{Energy spectra for $p$  and $Fe$ (upper) $Light$  and
$Heavy$ (lower) components, and their sum.
\label{fig:spettrie}}
\end{center}
\end{figure}
These resulting spectra for $p$-$Fe$ and $Light$-$Heavy$ components, 
further corrected for the 
detection efficiency of the surface array,
are reported in Fig. \ref{fig:spettrie} together with their sum.
Within such procedure, for each bin in  $log(E)$ we calculate 
$<logA>$ from the expression:
\begin{equation}
<logA> = \frac{p_1 M_k^p Log(A^p) + p_2 w_k M_k^{Fe} Log(A^{Fe})}
{M_k^p + p_2 w_k M_k^{Fe} }
\end{equation}
where $M_k^p$ and $M_k^{Fe}$ are the numbers of simulated
events from $p$ and $Fe$ contributing to the k-th energy bin. The same 
expression, with obvious modifications, is used for the $Light$-$Heavy$
composition.
The results are depicted in Figs.~\ref{fig:aval},
the two curves delimiting the error band. 
The results show an increase in $<logA>$
through the energy corresponding to the knee position.

\section{Conclusions}
The analysis of the N$_e$-N$_\mu$  ($E_{\mu} > 1.3$ TeV) data collected by
the MACRO/EAS-TOP collaboration at the Gran Sasso Laboratories
points to a primary composition becoming heavier around the
knee of the primary spectrum (in the energy region $10^{15}-10^{16}$ eV).
The result is in very good agreement with the measurements
of EAS-TOP alone at the surface, wich detects muons with energy
$E_{\mu} > 1$ GeV, using the same (QGSJET) interaction model
\cite{as6}.
GeV and TeV muons are produced in different kinematic regions:
in the central one and at the edges of the
fragmentation region respectively.
The present measurements show therefore a good consistency
of the interaction model in describing the yield of secondaries
over a wide rapidity region.

\end{document}